**Title: How to interpret hazard ratios**

**Jonathan W. Bartlett[1*], Dominic Magirr[2], Tim P. Morris[3]**

[1] Department of Medical Statistics, London School of Hygiene & Tropical Medicine, UK

[2] Statistical Methodology, Novartis Pharmaceuticals AG, Basel, Switzerland

[3] Statistical Methodology, Novartis Pharmaceuticals UK Ltd., London, UK

*Corresponding author. Department of Medical Statistics, London School of Hygiene & Tropical Medicine, Keppel Street, London, WC1E 7HT

**Abstract**

The hazard ratio, typically estimated using Cox's famous proportional hazards model, is the most common effect measure used to describe the association or effect of a covariate on a time-to-event outcome. In recent years the hazard ratio has been argued by some to lack a causal interpretation, even in randomised trials, and even if the proportional hazards assumption holds. This is concerning, not least due to the ubiquity of hazard ratios in analyses of time-to-event data. We review these criticisms, describe how we think hazard ratios should be interpreted, and argue that they retain a valid causal interpretation. Nevertheless, alternative measures may be preferable to describe effects of exposures or treatments on time-to-event outcomes.

**Key messages**

- Interpretation of hazards and hazard ratios comparing treatment or exposure groups is complicated by the possible effects of unmeasured (frailty) variables on event times.
- Hazard ratios estimated from randomised trials or from observational studies with sufficient confounder adjustment have a causal interpretation at a population level, but this interpretation should consider the possible effects of such unmeasured frailty variables.

- Other effect measures, such as contrasts of survival or event probabilities at key landmark times of interest may be preferable to hazard ratios for communicating treatment or exposure effects for time-to-event outcomes.

**Word count:** 2396

**Introduction**

Time-to-event outcomes are measured in many areas of health research. By far the most popular approach for analysing the relationships between time to event and covariates is Cox's famous *proportional hazards* model. Cox's model specifies how the hazard, the instantaneous failure rate among survivors at a given time, statistically depends on covariates. In a randomised trial, Cox's model with treatment group as the sole covariate specifies that the ratio of hazards between the two treatment groups remains constant over follow-up time – the proportional hazards assumption. When this assumption is deemed not to hold, researchers may instead partition follow-up and estimate hazard ratios (HRs) separately in each period of follow-up, which we refer to as period-specific HRs.

In recent years, some researchers have argued that HRs do not have a valid causal interpretation, even when the data arise from randomised trials, and even when the proportional hazards assumption holds [1]. In this article, we review the criticisms, articulate how we think HRs should be interpreted and, in so doing, argue that HRs do have a causal interpretation.

**Hazards, hazard ratios, and Cox's proportional hazards model**

Suppose we conduct a two-arm randomised trial. We let $Z$ denote the randomised treatment group for a randomly selected individual, with $Z = 0$ denoting randomisation to the control arm and $Z = 1$ denoting randomisation to the new treatment, or 'research' arm. The hazard (event rate) at time *t* in the control group is

$$h(t|Z=0) = \lim_{\delta t \to 0} \frac{P(t \leq T < t + \delta t | T \geq t, Z = 0)}{\delta t},$$

where $T$ denotes the event time of a randomly chosen individual. The hazard in the research arm is defined analogously. The (true) hazard ratio comparing the research group to the control group at a particular time *t* is thus

$$HR(t; Z) = \frac{h(t|Z=1)}{h(t|Z=0)}$$

Cox's proportional hazards model with treatment as covariate specifies that

$$h(t|Z) = h_0(t)e^{\beta Z}$$

where $h_0(t)$ denotes an unspecified 'baseline' hazard. The Cox model thus assumes that $HR(t; Z) = e^\beta$, independent of *t*, which is referred to as the proportional hazards assumption. When the proportional hazards assumption is not deemed plausible, a common approach is to partition follow-up into non-overlapping periods and fit an extended Cox model which estimates a distinct HR parameter comparing treatment groups in each period [2].

Under the proportional hazards assumption, one interpretation sometimes given to $e^\beta$, including in some textbooks [3], is that it is the factor by which an individual's hazard will be increased if they receive the research treatment rather than control. Such an interpretation as a common, individual-level effect is not generally justifiable due to the ubiquitous presence of so-called 'frailty' – unmeasured factors that influence individuals' hazards [1]. Frailty similarly affects the interpretation of period specific HRs [4, 5].

**The effects of frailty on hazard**

To understand the effects of frailty, suppose that there exists an as-yet-undiscovered genetic risk factor *X* which affects survival. For half of the population $X = 0$ and for the other half $X = 1$. Within groups defined by the genetic risk factor *X* and randomised treatment *Z*, time to death is exponentially distributed with constant hazard as specified by the values in Table 1.

Table 1: Hazard values $h(t|X,Z)$ by genetic factor X and treatment group Z in idealised randomised trial.

| Group | Hazard |
| --- | --- |
| $X = 1, Z = 0$ (control) | 0.5 |
| $X = 1, Z = 1$ (research) | 0.25 |
| $X = 0, Z = 0$ (control) | 0.1 |
| $X = 0, Z = 1$ (research) | 0.05 |

Figures 1a and 1b show the resulting true survival functions in the control and research arms respectively, with 500 individuals randomised to each treatment arm. Each shows the survival functions $S(t|X,Z)$ for the $X = 0$ and $X = 1$ individuals within each treatment arm. Since X is unmeasured, Figures 1a and 1b also show what we can *actually* estimate: the treatment group specific overall survival function $S(t|Z)$, ignoring or marginal to X, represented by the thicker solid lines. These overall treatment arm-specific survival functions $S(t|Z)$ are weighted averages of the respective $X = 0$ and $X = 1$ survival functions $S(t|X = 0, Z)$ and $S(t|X = 1, Z)$, with weights corresponding to the proportion without and with the genetic risk factor, here 0.5.

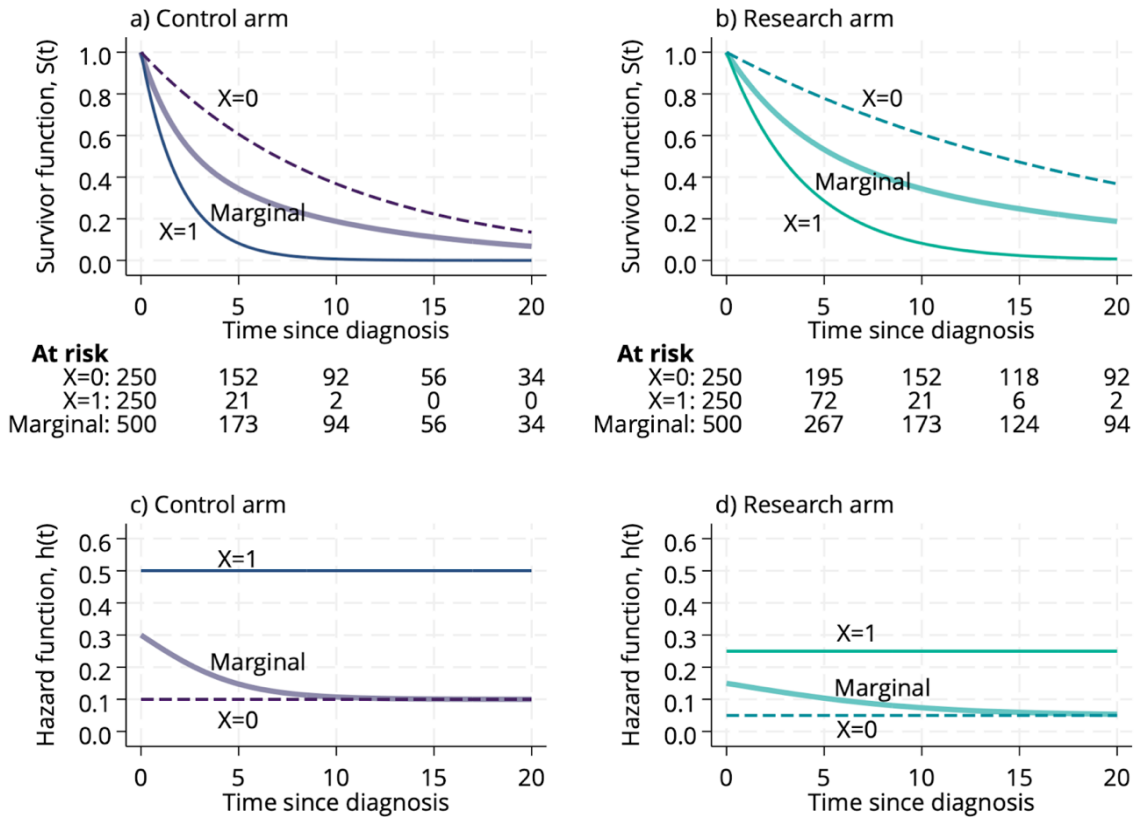

*Figure 1: Survival and hazard functions by treatment group and unmeasured genetic risk factor X in idealised trial.*

Figures 1c and 1d show the corresponding hazard functions. In the control arm, because initially there are equal proportions of $X = 0$ and $X = 1$ individuals, the hazard marginal to X, $h(t|Z = 0)$, is the average of the X-specific hazards 0.1 and 0.5. As time passes, the hazard $h(t|Z = 0)$ reduces, despite the hazard in groups defined by the genetic factor (i.e. $h(t|X,Z)$) remaining constant over time. This is because, as Figure 1a shows, at later follow-up times, the survivors are dominated by $X = 0$ individuals. Consequently, the hazard marginal to X, $h(t|Z = 0)$, increasingly resembles the hazard in the $X = 0$ group. A similar phenomenon occurs in the research arm. This simple illustration shows that one should not generally interpret a hazard as a common hazard pertaining to each individual in the study, but rather as the instantaneous event rate in the group of survivors at each time.

Now consider HRs comparing the research arm to the control arm. Recall from Table 1 that among individuals without the genetic risk factor ($X = 0$), the (time-constant) HR comparing the research arm to the control arm is $\frac{h(t|X=0,Z=1)}{h(t|X=0,Z=0)} = \frac{0.05}{0.1} = 0.5$, and among individuals with the genetic risk factor ($X = 1$), the (time-constant) HR comparing the research arm to the control arm is $\frac{h(t|X=1,Z=1)}{h(t|X=1,Z=0)} = \frac{0.25}{0.5} = 0.5$. Thus, conditional on or stratified by the risk factor $X$, there is a common, time-constant HR of 0.5. Recalling that $X$ is unmeasured, the HR we can estimate is marginal to $X$. The true HR (marginal to $X$) comparing research to control arms at a given time, $HR(t; Z)$, is the ratio of the true marginal hazards $h(t|Z)$ in each arm, and is shown in Figure 2.

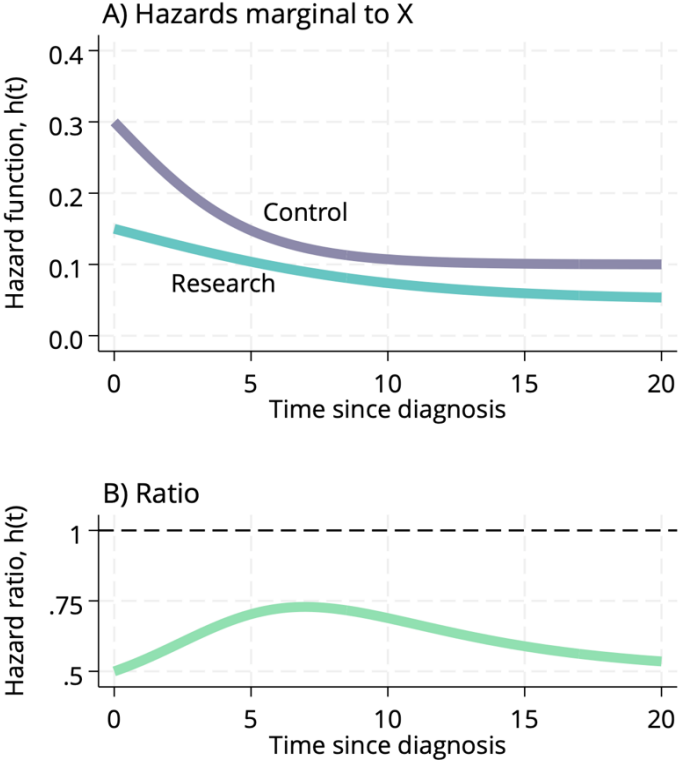

*Figure 2: Hazards in each treatment arm, marginal to the genetic risk factor X, and their ratio.*

Figure 2 shows $HR(t; Z)$ is close to 0.5 early in the follow-up. This is because initially there is a 50:50 mix of $X = 0$ and $X = 1$ individuals among the survivors in both arms. As time progresses however, $HR(t; Z)$ begins to attenuate towards one, apparently indicating an attenuation of the treatment effect (as measured by HR) on hazard of death. This is because in the middle stages of follow-up, among the surviving individuals in the research arm, a larger proportion have $X = 1$ than among surviving individuals in the control arm. Since individuals with $X = 1$ have higher hazard, this leads to a time-specific HR closer to 1.

Eventually though, the survivors in the two treatment arms only consist of $X = 0$ individuals, and so $HR(t; Z)$ converges to the ratio of the $X = 0$ specific hazards $\frac{h(t|X=0,Z=1)}{h(t|X=0,Z=0)} = \frac{0.05}{0.1} = 0.5$. In our simplified setup, this *X*-specific HR comparing treatments is the same for $X = 0$ and $X = 1$ groups, but of course in reality this may not be the case.

The value of $HR(t; Z)$ is thus in general a complicated composition of the two conditional (on *X*) hazard ratios for treatment, together with the dynamically changing proportion of patients from each strata in the risk set. Because the risk sets being compared at a given time *t* are not (in general) similar in respect of baseline characteristics (e.g. *X* in our simplified example), hazard ratios have been argued to not represent a valid causal effect of treatment, even in a randomised trial [1, 6, 7].

**How to interpret hazard ratios**

As noted earlier, if $HR(t; Z)$ is observed to vary over time, a common approach is to split follow-up into intervals and estimate separate period specific HRs [2]. Suppose that in such an analysis, $HR(t; Z)$ is (approximately) constant within the specified periods. In this case, the resulting period specific HR estimates have the following interpretation: if the population is assigned the research treatment, the hazard or instantaneous rate of failure at times *t* within the corresponding period among survivors would be $HR(t; Z)$ times what it would have been at time *t* had the population instead been assigned control treatment. Crucially, the subset of the population which would survive to time *t* under research treatment is in general systematically different in respect of baseline risk factors (i.e. variables such as *X*) for outcome compared to the subset of the population which would survive to time *t* under control treatment. This is a direct consequence of the earlier effects of the treatments and frailty factors on survival.

If instead $HR(t; Z)$ is (approximately) constant over time, a Cox model including treatment arm would typically be fitted. In this case the interpretation given above continues to hold, with now the common HR value applying for all times *t*. We note that, even in this case, frailty factors can still exist in such a way that the HRs conditional on these variables (e.g. $\frac{h(t|X=x,Z=1)}{h(t|X=x,Z=0)}$) vary with time [6]. As such, $HR(t; Z)$ being constant over time does not imply the absence of such frailty factors and hence their impact on the correct interpretation of $HR(t; Z)$.

When in truth $HR(t; Z)$ varies over time, as in Figure 2, fitting a Cox model assuming $HR(t; Z)$ to be constant estimates a weighted average value. Unfortunately however, the true value of this quantity depends on the censoring distribution. While various methods have been developed to estimate alternative average hazard ratios whose true value is not affected by the censoring distribution [8], these are not often used in practice.

**Are hazard ratios causal?**

As noted earlier, hazard ratios have been argued to not represent a valid causal effect of treatment, even in a randomised trial [1, 6, 7]. Our descriptions earlier of how to interpret HRs at one level suggest they are – the value of $HR(t; Z)$ quantifies by how much the population hazard at time *t* would be modified if we treated the population with the research treatment rather than the control treatment.

To be more formal, we consider the definition of what constitutes a causal effect given by Hernán and Robins in their leading textbook on causal inference [9]. Suppose we are interested in estimating the effect of a treatment on a certain population of size *N*. Specifically, we consider individuals' potential event times $T_1(1), ... T_N(1)$ under the research arm treatment; that is, the survival times of the *N* patients had they been assigned to the research arm. Similarly, we have potential event times $T_1(0), ... T_N(0)$ under the control treatment.

An *individual causal effect* can be defined as a contrast of $T_i(1)$ and $T_i(0)$ for an individual *i*, for example $T_i(1) - T_i(0)$ or $T_i(1)/T_i(0)$. It is clear that a hazard ratio is not an individual causal effect. The same is true for most summary measures used in clinical trials, e.g. a mean difference or risk ratio, except under assumptions that would usually not be plausible (e.g. for a mean difference unless all individual level effects are identical).

A *population causal effect* can be defined as a contrast of $T_1(1), ... T_N(1)$ and $T_1(0), ... T_N(0)$. More precisely, Hernán and Robins define a population causal effect as "*a contrast of any functional of the marginal distributions of counterfactual outcomes under different actions or treatment values*". Here, *functional* means a rule for mapping the potential outcomes under a given treatment, e.g. $T_1(1), ... T_N(1)$, into a single number. So, for example, taking the median of $T_1(1), ... T_N(1)$ is applying the functional "median" to the distribution of the potential (or counterfactual) outcomes.

According to this definition, if $HR(t; Z)$ is constant over time and equal to $e^\beta$, $e^\beta$ is a population causal effect. To see why, recall that under proportional hazards $S(t|Z=1) = S(t|Z=0)^{e^\beta}$. Taking logs this means that $e^\beta = \frac{\log S(t|Z=1)}{\log S(t|Z=0)}$, that is, the ratio of log survival probabilities to any time *t* under the two treatments, which is a function of the two potential outcome distributions, since treatment is randomly assigned [7, 10]. Since $S(t|Z) = e^{-H(t|Z)}$, where $H(t|Z) = \int_0^t h(u|Z)du$ denotes the cumulative hazard function, $e^\beta$ is also a ratio of cumulative hazards [7].

When $HR(t; Z)$ is not constant over time, it is still equal to a contrast of functionals of the potential outcome distributions. To see this, recall that the hazard $h(t|Z = z)$ can be expressed as $\frac{f(t|Z=z)}{S(t|Z=z)}$, where $f(t|Z = z)$ denotes the density function of survival times in treatment group *z*, and hence it itself a functional of the potential outcomes under treatment *z*. It then follows that $HR(t; Z)$ as the ratio of $h(t|Z = 1)$ and $h(t|Z = 0)$ is also a functional of the two potential outcome distributions. In conclusion, the number $HR(t; Z)$ is a causal quantity according to Hernan and Robins' definition. As described earlier, its interpretation is complicated in general by the effects of frailty factors [10].

**Discussion**

Hazards and contrasts of them are fundamental quantities in the analysis of time-to-event data and, rightly or wrongly, analyses of time-to-event outcomes are still almost always based on Cox's proportional hazards model. Causal inference ideas, beginning from Hernán's influential 'Hazard of Hazard Ratios' paper, have led to a fuller understanding of the complexities involved in the interpretation of HRs [4]. In light of criticisms of the HR from a causal perspective, our aim here was to elucidate the key aspects of these criticisms in a simple example and to describe how we think HRs can be legitimately interpreted in a causal sense.

Our example was deliberately contrived to discuss and visualize HRs: the event was common and the effect of frailty immodest (conditional HR of 5). As has been noted by others previously, the impacts are smaller when the event is rare or the frailty / unmeasured variables have modest effects [6, 11].

Randomised trial analyses using Cox models often adjust for a small number of baseline covariates to improve statistical power, while analyses of non-randomised studies may do so to adjust for confounding. If the model used is correctly specified, the resultant HRs for treatment can be interpreted as we have described, conditional on the included covariates [11].

Given the complexities involved in correctly interpreting hazards and HRs, particularly when the HR is not constant over time, it may be preferable to quantify effects using alternative measures [11]. These include differences or ratios of survival or risk probabilities at clinically relevant landmark times, or contrasts of restricted mean survival to a given time horizon. These measures, in particular the differences or ratios of survival or risk probabilities, are easier to (correctly) interpret than the HR, can be defined without reference to modelling assumptions, and do not suffer from non-collapsibility [12]. They nevertheless require choice of a landmark or horizon time. In settings where the proportional hazards assumption is violated, often no single number will be able to adequately summarise the differences in survival curves, and display of the survival curves themselves and multiple effect measures may be needed to most clearly communicate the differences [12].

**Acknowledgements**

The authors are grateful for discussions on this topic with various colleagues over recent years, in particular Rhian Daniel and Stijn Vansteelandt.

**Author contributions**

Jonathan Bartlett conceived of the idea of the paper, and drafted the paper in conjunction with Tim Morris, including producing the figures. Dominic Magirr subsequently contributed to editing and writing the paper, in particular regarding the causal validity of HRs.

**Conflicts of interest**

Jonathan Bartlett's past and present institutions have received consultancy fees for his advice on statistical methodology from AstraZeneca, Bayer, Novartis and Roche. Jonathan has received consultancy fees from Bayer and Roche. Dominic Magirr owns shares in Novartis Pharma AG. Tim Morris owns shares in Novartis Pharma AG. He has received consultancy fees from Novartis Pharma AG, Bayer Healthcare Pharmaceuticals, Alliance Pharmaceuticals, Gilead Sciences and Kite Pharma.

**Supplementary material**

Derivations of the formulas used to create the plots shown are available in Supplementary material. Stata code for producing the plots is available at github.com/tpmorris/interpret_hr

**Funding acknowledgment**

Jonathan Bartlett was supported by the UK Medical Research Council (grants MR/T023953/1 and MR/T023953/2).

**Supplementary material**

We derive expressions for the survival and hazard marginal to the frailty variable *X* in a given treatment group. We omit conditioning on the treatment group variable *Z* here to simplify notation. Assume that event times are exponentially distributed in the $X = 0$ group with hazard $\lambda_0$ and in the $X = 1$ group with hazard $\lambda_1$. Then the corresponding survival functions are given by

$$S(t|X = 0) = e^{-\lambda_0 t}$$

$$S(t|X = 0) = e^{-\lambda_1 t}$$

The survival function marginal to *X* is then given by

$$S(t) = P(T > t) = P(X = 0)P(T > t|X = 0) + P(X = 1)P(T > t|X = 1)$$

$$= P(X = 0)e^{-\lambda_0 t} + P(X = 1)e^{-\lambda_1 t}$$

We now use the relationship between the hazard and the survival function, specifically that $h(t) = -\frac{d}{dt} \log S(t)$ to find the marginal hazard function. We have

$$\log S(t) = \log\{P(X = 0)e^{-\lambda_0 t} + P(X = 1)e^{-\lambda_1 t}\}$$

and so the marginal hazard is given by

$$h(t) = \frac{\lambda_0 P(X = 0)e^{-\lambda_0 t} + \lambda_1 P(X = 1)e^{-\lambda_1 t}}{S(t)}$$